\documentclass[aip,jcp,reprint]{revtex4-1}

\usepackage{graphicx}
\usepackage{dcolumn}
\usepackage{bm}
\usepackage{hyperref}
\usepackage[mathlines]{lineno}
\usepackage{color}
\usepackage[normalem]{ulem}
\usepackage{rotating}
\def\ph2{$p$-H$_2$}
\def\od2{{\it o}-D$_2$}
\def\oh2{{\it o}-H$_2$}
\def\he4{$^4$He}
\begin{document}


\title{Exchange-driven self-diffusion of nanoscale crystalline parahydrogen clusters on graphite}

\author{K. M. Kolevski}
\author {M. Boninsegni}
\email{m.boninsegni@ualberta.ca}
\affiliation{Department of Physics, University of Alberta, Edmonton, Alberta, Canada T6G 2H5}

\date{\today}

\begin{abstract}
Computer simulations yield evidence of superfluid behavior of nanoscale size clusters of parahydrogen adsorbed on a graphite substrate at low temperature ($T\lesssim 0.25$ K). Clusters with a number of molecules between 7 and 12 display concurrent superfluidity and crystalline order, reflecting the corrugation of the substrate. Remarkably, it is found that specific clusters with a number of molecules ranging between 7 and 12 self-diffuse on the surface like free particles, despite the strong pinning effect of the substrate. This effect is underlain by coordinated quantum-mechanical exchanges of groups of identical molecules, i.e., it has no classical counterpart.

\end{abstract}
\maketitle
\section{\label{Introduction}Introduction}

Quantum clusters, namely few-body systems whose physical behavior is dominated by quantum effects, occupy a unique position in the investigation of condensed and nuclear matter. Generally speaking, they provide unique insight into the emergence of collective properties of large assemblies as the cluster size is increased.
For example, clusters of parahydrogen (\ph2) have been the subject of much theoretical \cite{Sindzingre1991,Scharf1992,Kwon2002,Mezzacapo2006,Mezzacapo2007,Khairallah2007,Mezzacapo2007b,Li2010,Li2019,Boninsegni2020} and experimental \cite{Grebenev2000,Tejeda2004,Vilesov2008,Raston2012} work, spanning now a period of a few decades. The interest in such systems is spurred by a confluence of factors. First and foremost, liquid  \ph2 has long been regarded as a potential superfluid \cite{Ginzburg1972}. However, bulk \ph2 displays a strong tendency to crystallize, thereby preempting any superfluid transition, even in reduced dimensions \cite{Boninsegni2004,Boninsegni2013,Boninsegni2018}. On the other hand, some experimental evidence has been produced of  (theoretically predicted) superfluid behavior of nanoscale size \ph2 clusters of $\sim 15$ molecules \cite{Grebenev2000}.
\\ \indent
It is not immediately obvious what the effect of reducing the dimensionality of such a system would be. On the one hand, a lower coordination number suppresses the tendency of \ph2 clusters to crystallize. On the other hand, quantum-mechanical exchanges which play a crucial role in the stabilization of superfluid  phases also become less frequent. It has been shown numerically that small \ph2 clusters of up to 25 molecules in two dimensions (2D) at low temperature ($T=0.25$ K) display a strong superfluid response while at the same time maintaining a clearly identifiable solid-like structure. Thus, they can be regarded as nanoscale ``supersolids'' \cite{Idowu2014}. These predictions could potentially be verified experimentally by adsorbing \ph2 onto an attractive substrate strong enough to confine the molecules to quasi-2D while also weak enough to make the effect of corrugation negligible. Alkali metal substrates seem a good candidate for the observation of this kind of physics \cite{Chizmeshya1998}.
\\ \indent
The opposite limit is that of a strongly attractive, highly corrugated substrate, such as graphite. In this case, the strength of the substrate renders adsorbed clusters essentially 2D, i.e., with no ``beading up'' of molecules. A recent study of quasi-2D clusters of $^4$He adsorbed on graphite has yielded evidence of an intriguing interplay of superfluidity and atomic localization for clusters up to $\sim 30$ atoms \cite{Kolevski2025}. In the case of \ph2, which has a stronger propensity than $^4$He to crystallize, substrate corrugation may render the system ``classical'' for all practical purposes by strongly suppressing quantum-mechanical exchanges. Indeed, adsorbed monolayer films of virtually any atomic or molecular species on graphite, including the highly quantal helium and parahydrogen, display largely classical behavior, with possibly a transition from commensurate to incommensurate crystal occurring on varying coverage \cite{Nho2002,Hu2024}. The question remains, however, of whether interesting quantum-mechanical effects might take place in sufficiently small \ph2 clusters adsorbed on graphite. 
\\ \indent
Previous theoretical calculations for finite \ph2 clusters adsorbed on substrates have focused on the adsorption on the surface of fullerenes \cite{Turnbull2005}, yielding evidence of a commensurate-to-incommensurate transition as a function of coverage in the low temperature limit. In these systems, exchanges of \ph2 molecules are essentially non-existent, and the physics of the incommensurate layer is essentially the same as one would predict assuming a smooth spherical surface \cite{Hernandez2003}. 
\\ \indent
We present here results of an extensive theoretical study of (\ph2)$_N$ clusters comprising up to $N=16$ molecules, adsorbed on a graphite substrate, down to temperatures as low as 30 mK. We make use of state-of-the-art computational methodology, and base our calculations on a quantitatively reliable microscopic model of the system, accounting for the attractive strength and corrugation of the substrate. At temperature $T\lesssim 0.25$ K, all clusters stay together, i.e., there is no evidence of \ph2 evaporation, either off the substrate or over it. Quantum-mechanical exchanges of identical molecules occur with high frequency in almost all clusters with $N < 16$, with just one exception, namely (\ph2)$_7$, for which exchanges are strongly suppressed. 
\\ \indent
Clusters with $N > 6$ display a clear, solid-like structure, with molecules sitting at preferential graphite adsorption sites; this is in concomitance with quantum exchanges,  i.e., superfluidity. Qualitatively, these results are reminiscent of what is predicted for $^4$He clusters \cite{Kolevski2025}.  A surprising finding of this study is that some superfluid and solid-like \ph2 clusters diffuse throughout the substrate essentially as individual particles, all while retaining a preferred, orderly arrangement of molecules. This effect, to our knowledge never observed nor predicted for any other atomic or molecular clusters, is crucially underlain by quantum-mechanical exchanges of indistinguishable molecules, which allow clusters to execute what ultimately amounts to rigid translational motion over the substrate. We opine that such an intriguing phenomenon should lend itself to experimental observation.

\section{\label{Methodology}Methodology}

We make use of an accepted microscopic model of the system, utilized in essentially all previous comparable studies, in which hydrogen molecules are considered as point-like particles of spin zero (hence obeying Bose statistics). We consider finite \ph2 clusters consisting of up to 16 molecules adsorbed on a graphite substrate situated on the basal plane at $z=0$. The system is enclosed in a cuboid simulation cell with periodic boundary conditions in all directions (although the dimensions are chosen large enough to render the boundary conditions immaterial).

The quantum-mechanical many-body Hamiltonian of the system reads as follows:
\begin{equation}
    \hat H = -\lambda\sum_{i}\nabla^2_{i}+\sum_{i<j}v(r_{ij})+\sum_{i}V({\mathbf r}_{i}).
\end{equation}
Here, $\lambda=12.031$ K\AA$^{2}$, 
the first and third sums run over all $N$ \ph2 molecules; the second sum runs over all pairs of molecules, with $r_{ij}\equiv |{\mathbf r}_i-{\mathbf r}_j|$, ${\mathbf r}_i$ being the position of the $i$th molecule.
We describe the interaction among hydrogen molecules ($v$) using the standard Silvera-Goldman intermolecular pair potential \cite{Silvera1978}, while for the interaction ($V$) of a hydrogen molecule with the graphite substrate we utilize the generalization of the Carlos-Cole potential \cite{Carlos1980} for a helium atom in the vicinity of a graphite substrate proposed in Ref. \cite{Nho2002}.
\\ \indent
Our computational methodology is the canonical variant \cite{Mezzacapo2006,Mezzacapo2007} of the continuous-space Worm Algorithm \cite{Boninsegni2006,Boninsegni2006b}, a robust technique based on Feynman's space-time approach to quantum statistical mechanics \cite{Feynman1965}, allowing one to compute numerically exact estimates of thermodynamic observables for Bose systems at a finite temperature $T=1/\beta$ (we set the Boltzmann constant $k_B=1$). Technical details of the simulations not explicitly discussed here are identical with those of Ref. \cite{Hu2024}.


The main quantities of interest in this study, besides energetics and structure of the clusters, are those that directly characterize the superfluid response (i.e., the frequency and length of exchanges) and the cluster self-diffusion over the surface of the graphite substrate. The latter is described by the center-of-mass imaginary-time diffusion coefficient
\begin{equation}\label{einstein}
D(\tau) = \frac{\langle[R_{CM}(\tau)-R_{CM}(0)]^2\rangle}{6\Lambda\tau}
\end{equation}
where $\langle ...\rangle$ stands for thermal average, $\Lambda\equiv\lambda/N$ and 
\begin{equation}
R_{CM}(\tau)=\frac{1}{N}\sum_i{\mathbf r}_i(\tau)
\end{equation}
is the position of the center of mass of the cluster at imaginary time $\tau$, with $0\le \tau\le \beta$. Because the system considered here is essentially 2D, $D(\tau)$ is dominated by molecular motion along the substrate.  In simulations of bulk systems with periodic boundary conditions, $D(\beta)$ reduces to the well-known {\it winding number estimator} of the superfluid fraction \cite{Pollock1987}. 

In the limit of vanishing density (i.e., for a finite system)  $D(\beta)=0$ at any finite temperature, as the positions of all the particles at $\tau=\beta$ must be the same as those at $\tau=0$, modulo a permutation of particle labels, and the superfluid response must be obtained through a different estimator \cite{Sindzingre1989}.
However, $D(\tau)$ can also be meaningfully utilized for a finite system, specifically to distinguish a situation in which individual particles, or small assemblies thereof, enjoy relative mobility or are instead ``pinned'' at specific (e.g., lattice) locations.

In order to illustrate this point, consider first the case of a single free particle in three dimensions, for which $D(\tau)$ can be computed analytically, i.e., \begin{equation}D(\tau)= D_F(\tau)\equiv 1-\frac{\tau}{\beta}.\end{equation} 
Interactions, either with other particles or with an external potential, generally limit the amount by which a particle diffuses in a given imaginary time interval (i.e., the spatial extension of its world line); in case of localization (e.g., in a crystal), $D(\tau)$ is a rapidly decaying function, whereas if the particle remains delocalized, the ratio 
\begin{equation}\label{delta}\delta(\tau)\equiv \frac{D_F(\tau)}{D(\tau)}\end{equation} 
plateaus to a finite value as $\tau\to\beta/2$ (for $\beta\to\infty$), which one can use to define an {\it effective mass} $(m^\star/m)= \delta(\beta/2) $ ($m$ is the bare particle mass) \cite{Boninsegni1995}. On the other hand, monotonic growth of $\delta(\tau)$   signals particle localization (i.e., ``infinite effective mass''). 

As shown below, the quantity $\delta(\tau)$ defined through Eq. (\ref{einstein}) for the center-of-mass of a (\ph2)$_N$ cluster  adsorbed on graphite, provides cogent information regarding the mobility of the cluster on the substrate. In particular, it allows one to connect the surprising mobility of specific solid-like clusters to exchanges, often involving many or sometimes even all of the molecules in a cluster, permitting them to move across the substrate while retaining their geometrically ordered structure.
\begin{figure}[!t]
    \includegraphics[width=\linewidth]{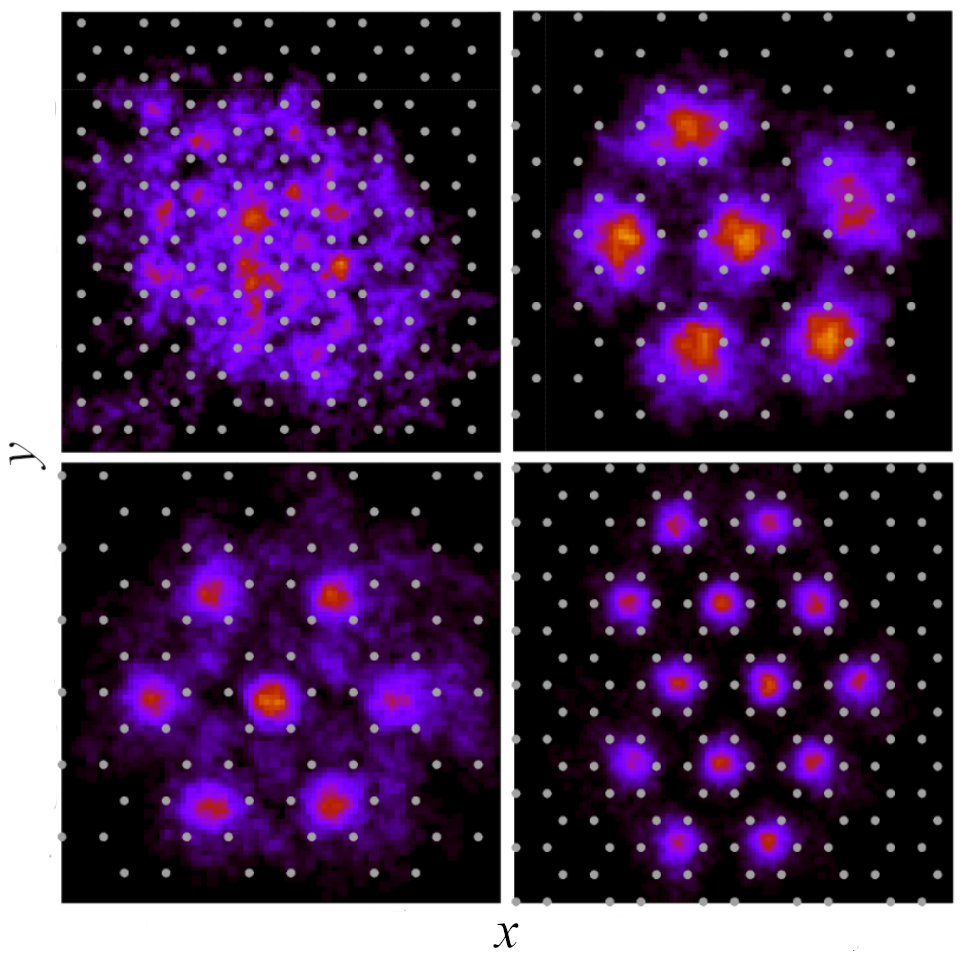}
    \caption{Molecular density maps (obtained from particle world lines) for representative configurations of a (\ph2)$_N$ cluster adsorbed on graphite at a temperature $T=0.125$ K. Maps shown are for $N=6$ (top left), $N=6$ {\it without} quantum exchanges (top right), $N=7$ (bottom left), and $N=13$ (bottom right). Dots arranged on a hexagonal lattice represent the top layer of carbon atoms on the graphite substrate (nearest-neighbor distance is 1.42 \AA.) }
    \label{fig:heatmap1}
\end{figure}
\begin{figure}[!h]
    \includegraphics[width=\linewidth]{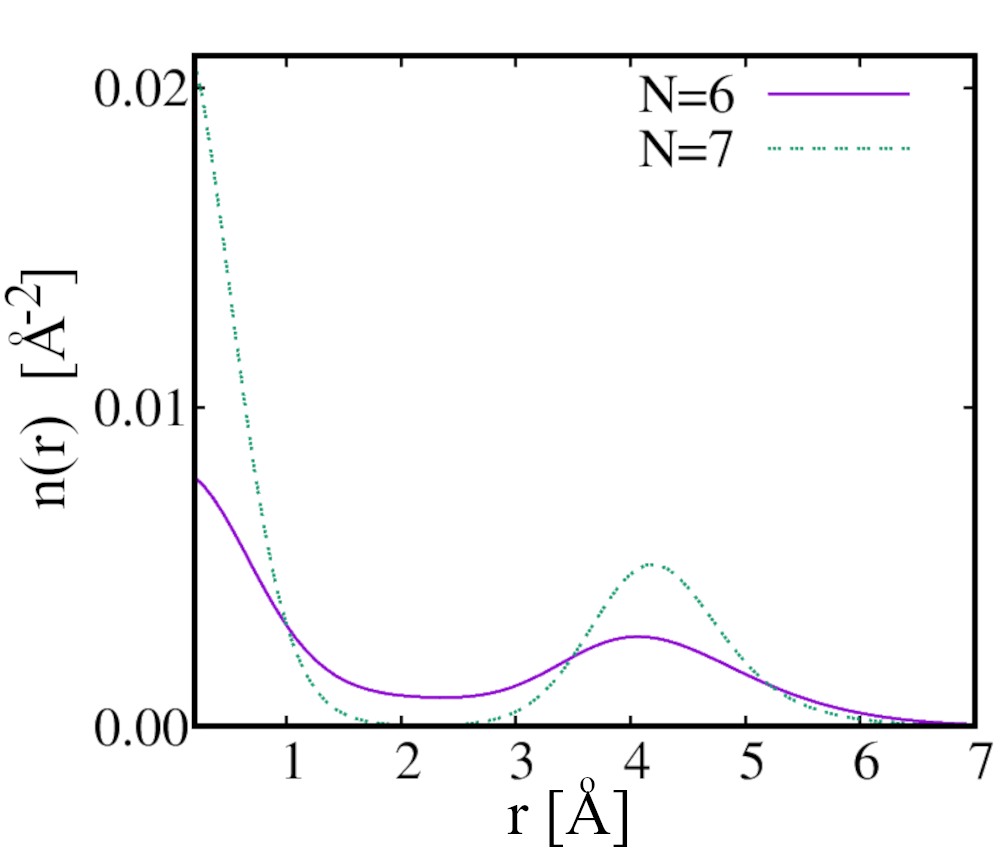}
    \caption{Radially averaged 2D particle density for the $N=6$ and $N=7$ clusters as a function of distance from the center of mass. For $N=6$, the consistent density of molecules throughout the cluster indicates a fluid-like lack of structure. For $N=7$, the high-peaked oscillations separated by a region of zero density indicate an orderly solid-like structure.}
    \label{fig:radialdensity}
\end{figure}

\section{\label{Results and Discussion}Results and Discussion}

Fig. \ref{fig:heatmap1} shows density maps computed from imaginary-time world lines for clusters comprising $N=6, 7,$ and $13$ \ph2 molecules at a temperature $T=0.125$ K. These images correspond to instantaneous configurations generated in the course of long simulations, but can be regarded as qualitatively representative, and visually illustrative, of the physics of the underlying systems. 
\\ \indent
The first aspect to discuss is the importance of quantum-mechanical exchanges of identical molecules in shaping the physical behavior of these systems. In general, it is observed that quantum effects, for those clusters in which they manifest themselves most significantly, become important at temperatures $\lesssim 0.25$ K. 
Consider, for example, the (\ph2)$_6$ cluster, for which results are shown in the top row of Fig. \ref{fig:heatmap1}. Specifically, we compare results obtained with (left) and without (right) quantum exchanges; excluding exchanges is equivalent to regarding molecules as {\it distinguishable}.
A striking difference emerges between the two cases: neglecting exchanges pins the molecules at regular positions, while including exchanges delocalizes them, causing the cluster to lose its ordered structure and concomitantly develop a strong superfluid response, an effect absent for distinguishable molecules.

\begin{figure}[!t]
    \includegraphics[width=\linewidth]{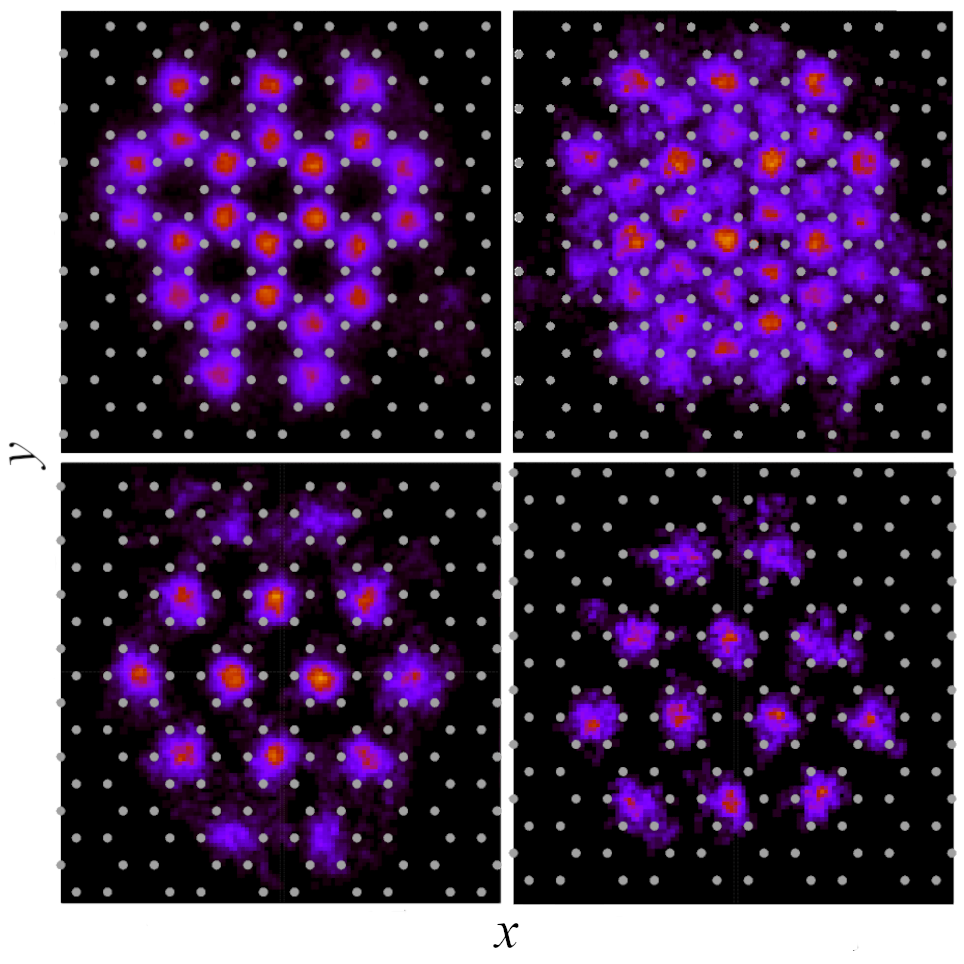}
    \caption{Molecular density maps (obtained from particle world lines) for representative configurations of a (\ph2)$_{12}$ cluster adsorbed on graphite at a temperature $T=0.0625$ K (top left), $T=0.125$ K (top right), $T=0.25$ K (bottom left), and $T=1$ K (bottom right). Dots arranged on a hexagonal lattice represent the top layer of carbon atoms on the graphite substrate (nearest-neighbor distance is 1.42 \AA.)}
    \label{fig:heatmap2}
\end{figure}

This is, of course, not a new observation. It has been long known that exchanges are crucial in stabilizing fluid phases against crystallization, not just in clusters but in bulk systems as well (e.g., helium) \cite{Mezzacapo2007,Boninsegni2012}. We find that clusters with $N \le 6$ are superfluid and structureless (see Fig. \ref{fig:radialdensity}), whereas for $N> 6$ an ordered arrangement of \ph2 molecules appears, as shown in the bottom part of Fig. \ref{fig:heatmap1} for $N=7$ and 13. Exchanges remain present, to different degrees, for clusters with $N < 13$; they are abruptly (and essentially entirely) suppressed for $N \ge 13$, where $P_{ex}$ becomes essentially zero within statistical noise, at least down to the lowest temperature for which simulations were carried out, namely 30 mK.  
\\ \indent
Qualitatively, these findings mimic those observed for $^4$He clusters on graphite \cite{Kolevski2025}. Indeed, just like for helium clusters, the solid-like arrangement of molecules locally mimics that of the commensurate ($C_{1/3}$) equilibrium monolayer structure \cite{Nho2002,Hu2024}.  An important difference, however, is that while exchanges are increasingly confined to the surface of a $^4$He cluster as its size is increased, in the case of \ph2 clusters they consistently involve all molecules in the cluster. This was already reported for \ph2 clusters in 2D, in the absence of a substrate \cite{Idowu2014}, and can be interpreted as the tendency of these clusters to act like nanoscale supersolids (with all the caveats required whenever one is talking about ``phases'' of finite systems of such a small size).

\begin{table}[!t]
    \centering
    \caption{Computed exchange frequency for (\ph2)$_N$ clusters at $T=0.125$ K. The estimated uncertainty on this quantity is $\lesssim 0.01$.}
    \label{tab:exchanges}
    \begin{tabular}{cc}
    \hline
    $N$ & $P_{ex}$ \\
    \hline
    $4$ & $0.51$ \\
    $5$ & $0.56$ \\
    $6$ & $0.54$ \\
    $7$ & $0.07$ \\
    $8$ & $0.43$ \\
    $9$ & $0.43$ \\
    $10$ & $0.41$ \\
    $11$ & $0.31$ \\
    $12$ & $0.43$ \\
    $13$ & $0.00$ \\
    \hline
    \end{tabular}
\end{table}

Quantitatively, the occurrence of quantum exchanges can be estimated by computing the probability ($P_{ex}$) for a \ph2 molecule to be involved in a cycle of exchange of arbitrary length (i.e., involving between 2 and $N$ molecules). Table \ref{tab:exchanges} shows results for $P_{ex}$ for all the clusters studied, at $T=0.125$ K.
We see that, for clusters of $N\leq12$ molecules the frequency of quantum exchanges is relatively high, hovering around $\sim 40\%$ for all clusters {\it except} for the $N=7$ one, for which it dips to roughly $\sim 7\%$. For clusters of $N=13$ molecules or more, the frequency of exchanges drops to almost zero. This suggests that the highly symmetrical arrangement of  molecules of the (\ph2)$_7$ cluster (shown in the bottom left panel of Fig. \ref{fig:heatmap1}), with six molecules forming a hexagon and one in the middle of it, renders exchanges energetically unfavorable. In this sense, this could be regarded as a ``magic'' cluster. Interestingly, this is not the case for this cluster in 2D, in the absence of a corrugated substrate, i.e., its observed stability in this work is a direct result of the substrate corrugation.
\\ \indent
In order to illustrate the effects of quantum exchanges as they progressively set on, as the temperature is lowered, we focus on the (\ph2)$_{12}$ cluster, for which representative density maps are shown in Fig. \ref{fig:heatmap2} at four different temperatures. At the highest temperature, namely $T=1$ K, the observed crystalline structure is that which minimizes the potential energy, i.e., the system behaves  essentially classically (aside of course from the significant kinetic energy of localization of the molecules). At these temperatures, exchanges of \ph2 molecules are virtually absent, i.e., molecules behave essentially as distinguishable particles.
\\ \indent
A remarkable effect begins to appear at $T=0.25$ K, becoming increasingly  more evident as the temperature is lowered even further, i.e., as quantum exchanges become frequent. Specifically, while crystalline order remains clear in every configuration, nevertheless in the $0\le\tau\le\beta$ imaginary-time interval ``shadows'' of molecules appear at crystalline sites other than those prevalently occupied. This is already visible at $T=0.25$ K (bottom left panel of Fig. \ref{fig:heatmap2}), where the configuration shows two horizontal adjacent sites at the top and bottom of the cluster occupied by molecules for a significant fraction of the imaginary time interval. Corresponding snapshots of the system at lower temperature, shown in the top panels of Fig. \ref{fig:heatmap2}, give the visual impression of a rigid spatial shift of the cluster as a whole. 
\\ \indent
\begin{figure}[!t]
    \includegraphics[width=\linewidth]{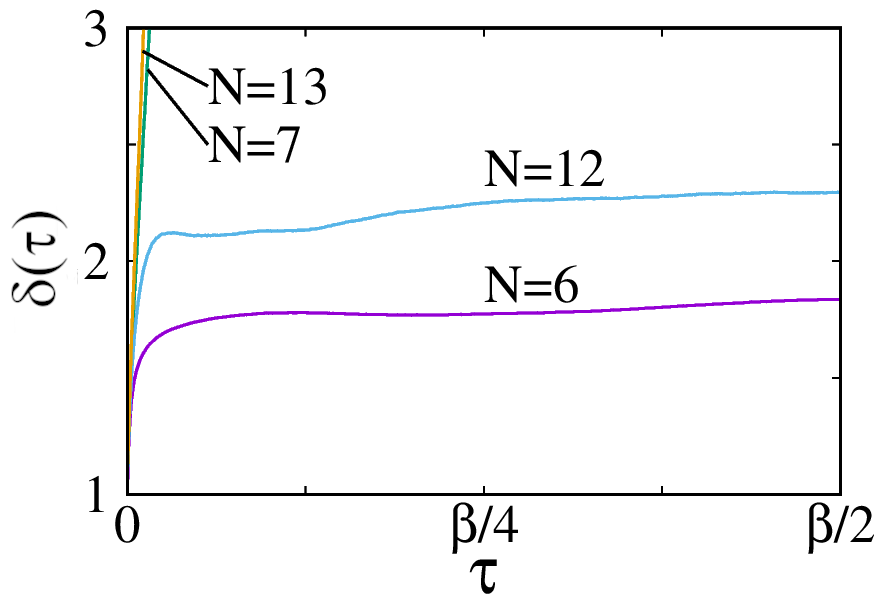}
    \caption{The quantity $\delta(\tau)$ defined in text for clusters of $N=6, 7, 12,$ and $13$ \ph2 molecules plotted against imaginary time at $T=0.125$ K.}
    \label{fig:selfdiffusion}
\end{figure}
One might plausibly think that what is observed in the various cases discussed here is merely the result of consecutive hops of individual molecules to unoccupied sites, with each molecule retracing its path in imaginary time, in order to return to its position at $\tau=0$. We contend, however, that this is not what actually happens. First, if exchanges are turned off (i.e., \ph2 molecules are considered distinguishable), the molecular configurations generated at the lowest temperature ($T=0.0625$ K) are essentially identical to those at $T=1$ K. In other words, configurations such as those shown in the top panel of Fig. \ref{fig:heatmap2} are {\it not} observed unless quantum-mechanical exchanges of \ph2 molecules are included in the simulations.  
Moreover, a careful visual examination of imaginary-time paths generated in the simulation shows that configurations that appear as a ``rigid'' shift of the whole cluster (e.g., top left panel of Fig. \ref{fig:heatmap2}) occur as a result of exchanges of one or more groups of molecules, and in some cases permutations involving {\it all} molecules in the cluster \footnote{An illustrative animation of representative imaginary-time many-particle paths for a (\ph2)$_8$ cluster at $T=0.125$ K can be viewed at \url{https://doi.org/10.5281/zenodo.16808901}.}.
\\ \indent
The physical observation is that exchanges of indistinguishable molecules impart to the cluster the ability to move around on the substrate in a way in which a single molecule cannot, diffusing though the corrugation, all while retaining its crystalline structure. The density maps in Figs. \ref{fig:heatmap1} and \ref{fig:heatmap2} are consistent with an instanton picture of the system, i.e., one in which imaginary-time paths feature \ph2 molecules mainly occupying specific spatial (lattice) locations, with scarcely any trace of their presence anywhere else, i.e., a typical ``hop'' from one lattice location to another occurs in a vanishingly small imaginary time.
\\ \indent
In order to characterize quantitatively the self-diffusion of clusters, we consider the quantity $\delta(\tau)$ defined in Eq. \ref{delta}. As mentioned above, two physically different regimes can be identified, one characterized by monotonic growth of $\delta(\tau)$, which corresponds to cluster localization, and the second in which $\delta(\tau)$ plateaus to a finite value as $\tau\to\beta/2$, which is indicative of self-diffusion.
Fig. \ref{fig:selfdiffusion} displays $\delta(\tau)$ for (\ph2)$_N$ clusters with $N=6, 7, 12,$ and $13$. Here, we see a clear difference between clusters with $N=7$ and $N=13$, which are localized, and those with $N=6$ and $N=12$ which self-diffuse over the graphite surface, essentially as if they were free particles. Clusters with $8\le N\le 12$ all feature the same behavior. Such self-diffusive ability crucially hinges on quantum exchanges, as we have verified in simulations in which \ph2 molecules are treated as distinguishable particles, yielding results for $\delta(\tau)$ similar to those for $N=7, 13$. Quantum exchanges are indeed strongly suppressed in clusters with $N=7$ and $N>12$ (see Table \ref{tab:exchanges}). Among those clusters that self-diffuse, there is a qualitative difference between clusters with $N\le 6$, which are featureless and liquid-like, and those with $8\le N \le 12$, which are instead simultaneously superfluid and crystalline.
\\ \indent
These results show that the physics of this system is qualitatively different from that of \he4 clusters adsorbed on the same substrate. (\he4)$_N$ clusters  with less than $N\sim20$ atoms are $\sim100\%$ superfluid and liquid-like, with no ``magic'' clusters. As the size of the cluster is increased, the $C_{1/3}$ commensurate crystalline structure \cite{Bretz1973,Nielsen1980,Greywall1993} nucleates in its center and the residual superfluid response is confined to a quasi-one-dimensional shell of atoms on the surface. This stands in stark contrast to the findings presented here, with a set of ``supersolid'' clusters (comprising between 8 and 12 \ph2 molecules) in which quantum exchanges set in at low temperature and involve all molecules. Additionally, these crystalline clusters display an unusual propensity to diffuse through the substrate corrugation, a phenomenon with no classical counterpart, which to our knowledge has not been previously reported or predicted.

\section{\label{Conclusion}Conclusion}

We have investigated the interplay between quantum exchanges and crystallization in (\ph2)$_N$ clusters adsorbed on a graphite substrate by means of numerically exact computer simulations. Our results demonstrate that, despite the presence of a highly corrugated substrate, quantum exchanges occur, and are crucially important in imparting remarkable and unusual physical properties to clusters of $8 \le N \leq 12$ \ph2 molecules. These clusters act like nanoscale ``supersolids'', simultaneously displaying crystalline arrangement of molecules and a strong propensity for quantum-mechanical exchanges. Such a behavior has already been predicted for small clusters of \ph2 in two and three dimensions; it is not believed to survive in bulk thermodynamic phases, a conclusion that generally applies to all many-body systems whose interparticle interactions feature a hard repulsive core at short distances \cite{Boninsegni2012b,Kora2019}. Supersolid clusters adsorbed on a graphite substrate display remarkable, unusual  center-of-mass mobility, i.e., self-diffusion and free-particle-like behavior. 
\\ \indent
Previous theoretical work on self-diffusion of Lennard-Jones type clusters on surfaces has been confined to the classical case \cite{Deltour1997, Ala-Nissila2002}. In this work, we show that self-diffusion of a {\it crystalline} cluster on a highly corrugated substrate can arise from quantum effects, specifically and most notably quantum-mechanical exchanges of indistinguishable particles. Thus, self-diffusion is a property of {\it supersolid} clusters, i.e., ones in which the regular arrangement of molecules is concomitant with quantum mechanical exchanges involving all molecules in the cluster. These findings might lend themselves to experimental observation as the investigation of hydrogen clusters on graphite has been extensively pursued using such techniques as scanning tunneling microscopy \cite{Hornekær2006}. 
\section*{Acknowledgments}
This work was supported by the Natural Sciences and Engineering Research Council of Canada under grant RGPIN 2024-05664.

\bibliography{references}

\begin{thebibliography}{43}%
\makeatletter
\providecommand \@ifxundefined [1]{%
 \@ifx{#1\undefined}
}%
\providecommand \@ifnum [1]{%
 \ifnum #1\expandafter \@firstoftwo
 \else \expandafter \@secondoftwo
 \fi
}%
\providecommand \@ifx [1]{%
 \ifx #1\expandafter \@firstoftwo
 \else \expandafter \@secondoftwo
 \fi
}%
\providecommand \natexlab [1]{#1}%
\providecommand \enquote  [1]{``#1''}%
\providecommand \bibnamefont  [1]{#1}%
\providecommand \bibfnamefont [1]{#1}%
\providecommand \citenamefont [1]{#1}%
\providecommand \href@noop [0]{\@secondoftwo}%
\providecommand \href [0]{\begingroup \@sanitize@url \@href}%
\providecommand \@href[1]{\@@startlink{#1}\@@href}%
\providecommand \@@href[1]{\endgroup#1\@@endlink}%
\providecommand \@sanitize@url [0]{\catcode `\\12\catcode `\$12\catcode `\&12\catcode `\#12\catcode `\^12\catcode `\_12\catcode `\%12\relax}%
\providecommand \@@startlink[1]{}%
\providecommand \@@endlink[0]{}%
\providecommand \url  [0]{\begingroup\@sanitize@url \@url }%
\providecommand \@url [1]{\endgroup\@href {#1}{\urlprefix }}%
\providecommand \urlprefix  [0]{URL }%
\providecommand \Eprint [0]{\href }%
\providecommand \doibase [0]{http://dx.doi.org/}%
\providecommand \selectlanguage [0]{\@gobble}%
\providecommand \bibinfo  [0]{\@secondoftwo}%
\providecommand \bibfield  [0]{\@secondoftwo}%
\providecommand \translation [1]{[#1]}%
\providecommand \BibitemOpen [0]{}%
\providecommand \bibitemStop [0]{}%
\providecommand \bibitemNoStop [0]{.\EOS\space}%
\providecommand \EOS [0]{\spacefactor3000\relax}%
\providecommand \BibitemShut  [1]{\csname bibitem#1\endcsname}%
\let\auto@bib@innerbib\@empty
\bibitem [{\citenamefont {Sindzingre}, \citenamefont {Ceperley},\ and\ \citenamefont {Klein}(1991)}]{Sindzingre1991}%
  \BibitemOpen
  \bibfield  {author} {\bibinfo {author} {\bibfnamefont {P.}~\bibnamefont {Sindzingre}}, \bibinfo {author} {\bibfnamefont {D.~M.}\ \bibnamefont {Ceperley}}, \ and\ \bibinfo {author} {\bibfnamefont {M.~L.}\ \bibnamefont {Klein}},\ }\href@noop {} {\bibfield  {journal} {\bibinfo  {journal} {Phys. Rev. Lett.}\ }\textbf {\bibinfo {volume} {67}},\ \bibinfo {pages} {1871} (\bibinfo {year} {1991})}\BibitemShut {NoStop}%
\bibitem [{\citenamefont {Scharf}, \citenamefont {Klein},\ and\ \citenamefont {Martyna}(1992)}]{Scharf1992}%
  \BibitemOpen
  \bibfield  {author} {\bibinfo {author} {\bibfnamefont {D.}~\bibnamefont {Scharf}}, \bibinfo {author} {\bibfnamefont {M.~L.}\ \bibnamefont {Klein}}, \ and\ \bibinfo {author} {\bibfnamefont {G.~J.}\ \bibnamefont {Martyna}},\ }\href {\doibase 10.1063/1.462994} {\bibfield  {journal} {\bibinfo  {journal} {J. Chem. Phys.}\ }\textbf {\bibinfo {volume} {97}},\ \bibinfo {pages} {3590} (\bibinfo {year} {1992})}\BibitemShut {NoStop}%
\bibitem [{\citenamefont {Kwon}\ and\ \citenamefont {Whaley}(2002)}]{Kwon2002}%
  \BibitemOpen
  \bibfield  {author} {\bibinfo {author} {\bibfnamefont {Y.}~\bibnamefont {Kwon}}\ and\ \bibinfo {author} {\bibfnamefont {K.~B.}\ \bibnamefont {Whaley}},\ }\href {\doibase 10.1103/PhysRevLett.89.273401} {\bibfield  {journal} {\bibinfo  {journal} {Phys. Rev. Lett.}\ }\textbf {\bibinfo {volume} {89}},\ \bibinfo {pages} {273401} (\bibinfo {year} {2002})}\BibitemShut {NoStop}%
\bibitem [{\citenamefont {Mezzacapo}\ and\ \citenamefont {Boninsegni}(2006)}]{Mezzacapo2006}%
  \BibitemOpen
  \bibfield  {author} {\bibinfo {author} {\bibfnamefont {F.}~\bibnamefont {Mezzacapo}}\ and\ \bibinfo {author} {\bibfnamefont {M.}~\bibnamefont {Boninsegni}},\ }\href {\doibase 10.1103/PhysRevLett.97.045301} {\bibfield  {journal} {\bibinfo  {journal} {Phys. Rev. Lett.}\ }\textbf {\bibinfo {volume} {97}},\ \bibinfo {pages} {045301} (\bibinfo {year} {2006})}\BibitemShut {NoStop}%
\bibitem [{\citenamefont {Mezzacapo}\ and\ \citenamefont {Boninsegni}(2007{\natexlab{a}})}]{Mezzacapo2007}%
  \BibitemOpen
  \bibfield  {author} {\bibinfo {author} {\bibfnamefont {F.}~\bibnamefont {Mezzacapo}}\ and\ \bibinfo {author} {\bibfnamefont {M.}~\bibnamefont {Boninsegni}},\ }\href {\doibase 10.1103/PhysRevA.75.033201} {\bibfield  {journal} {\bibinfo  {journal} {Phys. Rev. A}\ }\textbf {\bibinfo {volume} {75}},\ \bibinfo {pages} {033201} (\bibinfo {year} {2007}{\natexlab{a}})}\BibitemShut {NoStop}%
\bibitem [{\citenamefont {Khairallah}\ \emph {et~al.}(2007)\citenamefont {Khairallah}, \citenamefont {Sevryuk}, \citenamefont {Ceperley},\ and\ \citenamefont {Toennies}}]{Khairallah2007}%
  \BibitemOpen
  \bibfield  {author} {\bibinfo {author} {\bibfnamefont {S.~A.}\ \bibnamefont {Khairallah}}, \bibinfo {author} {\bibfnamefont {M.~B.}\ \bibnamefont {Sevryuk}}, \bibinfo {author} {\bibfnamefont {D.~M.}\ \bibnamefont {Ceperley}}, \ and\ \bibinfo {author} {\bibfnamefont {J.~P.}\ \bibnamefont {Toennies}},\ }\href {\doibase 10.1103/PhysRevLett.98.183401} {\bibfield  {journal} {\bibinfo  {journal} {Phys. Rev. Lett.}\ }\textbf {\bibinfo {volume} {98}},\ \bibinfo {pages} {183401} (\bibinfo {year} {2007})}\BibitemShut {NoStop}%
\bibitem [{\citenamefont {Mezzacapo}\ and\ \citenamefont {Boninsegni}(2007{\natexlab{b}})}]{Mezzacapo2007b}%
  \BibitemOpen
  \bibfield  {author} {\bibinfo {author} {\bibfnamefont {F.}~\bibnamefont {Mezzacapo}}\ and\ \bibinfo {author} {\bibfnamefont {M.}~\bibnamefont {Boninsegni}},\ }\href {\doibase 10.1103/PhysRevA.76.021201} {\bibfield  {journal} {\bibinfo  {journal} {Phys. Rev. A}\ }\textbf {\bibinfo {volume} {76}},\ \bibinfo {pages} {021201} (\bibinfo {year} {2007}{\natexlab{b}})}\BibitemShut {NoStop}%
\bibitem [{\citenamefont {Li}\ \emph {et~al.}(2010)\citenamefont {Li}, \citenamefont {Le~Roy}, \citenamefont {Roy},\ and\ \citenamefont {McKellar}}]{Li2010}%
  \BibitemOpen
  \bibfield  {author} {\bibinfo {author} {\bibfnamefont {H.}~\bibnamefont {Li}}, \bibinfo {author} {\bibfnamefont {R.~J.}\ \bibnamefont {Le~Roy}}, \bibinfo {author} {\bibfnamefont {P.-N.}\ \bibnamefont {Roy}}, \ and\ \bibinfo {author} {\bibfnamefont {A.~R.~W.}\ \bibnamefont {McKellar}},\ }\href {\doibase 10.1103/PhysRevLett.105.133401} {\bibfield  {journal} {\bibinfo  {journal} {Phys. Rev. Lett.}\ }\textbf {\bibinfo {volume} {105}},\ \bibinfo {pages} {133401} (\bibinfo {year} {2010})}\BibitemShut {NoStop}%
\bibitem [{\citenamefont {Li}\ \emph {et~al.}(2019)\citenamefont {Li}, \citenamefont {Zhang}, \citenamefont {Zeng}, \citenamefont {Le~Roy},\ and\ \citenamefont {Roy}}]{Li2019}%
  \BibitemOpen
  \bibfield  {author} {\bibinfo {author} {\bibfnamefont {H.}~\bibnamefont {Li}}, \bibinfo {author} {\bibfnamefont {X.-L.}\ \bibnamefont {Zhang}}, \bibinfo {author} {\bibfnamefont {T.}~\bibnamefont {Zeng}}, \bibinfo {author} {\bibfnamefont {R.~J.}\ \bibnamefont {Le~Roy}}, \ and\ \bibinfo {author} {\bibfnamefont {P.-N.}\ \bibnamefont {Roy}},\ }\href {\doibase 10.1103/PhysRevLett.123.093001} {\bibfield  {journal} {\bibinfo  {journal} {Phys. Rev. Lett.}\ }\textbf {\bibinfo {volume} {123}},\ \bibinfo {pages} {093001} (\bibinfo {year} {2019})}\BibitemShut {NoStop}%
\bibitem [{\citenamefont {Boninsegni}(2020)}]{Boninsegni2020}%
  \BibitemOpen
  \bibfield  {author} {\bibinfo {author} {\bibfnamefont {M.}~\bibnamefont {Boninsegni}},\ }\href {\doibase 10.1007/s10909-020-02493-4} {\bibfield  {journal} {\bibinfo  {journal} {J. Low Temp.Phys.}\ }\textbf {\bibinfo {volume} {201}},\ \bibinfo {pages} {193} (\bibinfo {year} {2020})}\BibitemShut {NoStop}%
\bibitem [{\citenamefont {Grebenev}\ \emph {et~al.}(2000)\citenamefont {Grebenev}, \citenamefont {Sartakov}, \citenamefont {Toennies},\ and\ \citenamefont {Vilesov}}]{Grebenev2000}%
  \BibitemOpen
  \bibfield  {author} {\bibinfo {author} {\bibfnamefont {S.}~\bibnamefont {Grebenev}}, \bibinfo {author} {\bibfnamefont {B.}~\bibnamefont {Sartakov}}, \bibinfo {author} {\bibfnamefont {J.}~\bibnamefont {Toennies}}, \ and\ \bibinfo {author} {\bibfnamefont {A.}~\bibnamefont {Vilesov}},\ }\href {\doibase 10.1126/science.289.5484.1532} {\bibfield  {journal} {\bibinfo  {journal} {Science}\ }\textbf {\bibinfo {volume} {289}},\ \bibinfo {pages} {1532} (\bibinfo {year} {2000})}\BibitemShut {NoStop}%
\bibitem [{\citenamefont {Tejeda}\ \emph {et~al.}(2004)\citenamefont {Tejeda}, \citenamefont {Fern\'andez}, \citenamefont {Montero}, \citenamefont {Blume},\ and\ \citenamefont {Toennies}}]{Tejeda2004}%
  \BibitemOpen
  \bibfield  {author} {\bibinfo {author} {\bibfnamefont {G.}~\bibnamefont {Tejeda}}, \bibinfo {author} {\bibfnamefont {J.~M.}\ \bibnamefont {Fern\'andez}}, \bibinfo {author} {\bibfnamefont {S.}~\bibnamefont {Montero}}, \bibinfo {author} {\bibfnamefont {D.}~\bibnamefont {Blume}}, \ and\ \bibinfo {author} {\bibfnamefont {J.~P.}\ \bibnamefont {Toennies}},\ }\href {\doibase 10.1103/PhysRevLett.92.223401} {\bibfield  {journal} {\bibinfo  {journal} {Phys. Rev. Lett.}\ }\textbf {\bibinfo {volume} {92}},\ \bibinfo {pages} {223401} (\bibinfo {year} {2004})}\BibitemShut {NoStop}%
\bibitem [{\citenamefont {Kuyanov-Prozument}\ and\ \citenamefont {Vilesov}(2008)}]{Vilesov2008}%
  \BibitemOpen
  \bibfield  {author} {\bibinfo {author} {\bibfnamefont {K.}~\bibnamefont {Kuyanov-Prozument}}\ and\ \bibinfo {author} {\bibfnamefont {A.~F.}\ \bibnamefont {Vilesov}},\ }\href {\doibase 10.1103/PhysRevLett.101.205301} {\bibfield  {journal} {\bibinfo  {journal} {Phys. Rev. Lett.}\ }\textbf {\bibinfo {volume} {101}},\ \bibinfo {pages} {205301} (\bibinfo {year} {2008})}\BibitemShut {NoStop}%
\bibitem [{\citenamefont {Raston}\ \emph {et~al.}(2012)\citenamefont {Raston}, \citenamefont {J\"ager}, \citenamefont {Li}, \citenamefont {Le~Roy},\ and\ \citenamefont {Roy}}]{Raston2012}%
  \BibitemOpen
  \bibfield  {author} {\bibinfo {author} {\bibfnamefont {P.~L.}\ \bibnamefont {Raston}}, \bibinfo {author} {\bibfnamefont {W.}~\bibnamefont {J\"ager}}, \bibinfo {author} {\bibfnamefont {H.}~\bibnamefont {Li}}, \bibinfo {author} {\bibfnamefont {R.~J.}\ \bibnamefont {Le~Roy}}, \ and\ \bibinfo {author} {\bibfnamefont {P.-N.}\ \bibnamefont {Roy}},\ }\href {\doibase 10.1103/PhysRevLett.108.253402} {\bibfield  {journal} {\bibinfo  {journal} {Phys. Rev. Lett.}\ }\textbf {\bibinfo {volume} {108}},\ \bibinfo {pages} {253402} (\bibinfo {year} {2012})}\BibitemShut {NoStop}%
\bibitem [{\citenamefont {Ginzburg}\ and\ \citenamefont {Sobyanin}(1972)}]{Ginzburg1972}%
  \BibitemOpen
  \bibfield  {author} {\bibinfo {author} {\bibfnamefont {V.~L.}\ \bibnamefont {Ginzburg}}\ and\ \bibinfo {author} {\bibfnamefont {A.~A.}\ \bibnamefont {Sobyanin}},\ }\href@noop {} {\bibfield  {journal} {\bibinfo  {journal} {JETP Lett.}\ }\textbf {\bibinfo {volume} {15}},\ \bibinfo {pages} {242} (\bibinfo {year} {1972})}\BibitemShut {NoStop}%
\bibitem [{\citenamefont {Boninsegni}(2004)}]{Boninsegni2004}%
  \BibitemOpen
  \bibfield  {author} {\bibinfo {author} {\bibfnamefont {M.}~\bibnamefont {Boninsegni}},\ }\href {\doibase 10.1103/PhysRevB.70.193411} {\bibfield  {journal} {\bibinfo  {journal} {Phys. Rev. B}\ }\textbf {\bibinfo {volume} {70}} (\bibinfo {year} {2004}),\ 10.1103/PhysRevB.70.193411}\BibitemShut {NoStop}%
\bibitem [{\citenamefont {Boninsegni}(2013)}]{Boninsegni2013}%
  \BibitemOpen
  \bibfield  {author} {\bibinfo {author} {\bibfnamefont {M.}~\bibnamefont {Boninsegni}},\ }\href {\doibase 10.1103/PhysRevLett.111.235303} {\bibfield  {journal} {\bibinfo  {journal} {Phys. Rev. Lett.}\ }\textbf {\bibinfo {volume} {111}} (\bibinfo {year} {2013}),\ 10.1103/PhysRevLett.111.235303}\BibitemShut {NoStop}%
\bibitem [{\citenamefont {Boninsegni}(2018)}]{Boninsegni2018}%
  \BibitemOpen
  \bibfield  {author} {\bibinfo {author} {\bibfnamefont {M.}~\bibnamefont {Boninsegni}},\ }\href {\doibase 10.1103/PhysRevB.97.054517} {\bibfield  {journal} {\bibinfo  {journal} {Phys. Rev. B}\ }\textbf {\bibinfo {volume} {97}},\ \bibinfo {pages} {054517} (\bibinfo {year} {2018})}\BibitemShut {NoStop}%
\bibitem [{\citenamefont {Idowu}\ and\ \citenamefont {Boninsegni}(2014)}]{Idowu2014}%
  \BibitemOpen
  \bibfield  {author} {\bibinfo {author} {\bibfnamefont {S.}~\bibnamefont {Idowu}}\ and\ \bibinfo {author} {\bibfnamefont {M.}~\bibnamefont {Boninsegni}},\ }\href {\doibase 10.1063/1.4878376} {\bibfield  {journal} {\bibinfo  {journal} {J. Chem. Phys.}\ }\textbf {\bibinfo {volume} {140}},\ \bibinfo {pages} {204310} (\bibinfo {year} {2014})}\BibitemShut {NoStop}%
\bibitem [{\citenamefont {Chizmeshya}, \citenamefont {Cole},\ and\ \citenamefont {Zaremba}(1998)}]{Chizmeshya1998}%
  \BibitemOpen
  \bibfield  {author} {\bibinfo {author} {\bibfnamefont {A.}~\bibnamefont {Chizmeshya}}, \bibinfo {author} {\bibfnamefont {M.}~\bibnamefont {Cole}}, \ and\ \bibinfo {author} {\bibfnamefont {E.}~\bibnamefont {Zaremba}},\ }\href {\doibase 10.1023/A:1022556227148} {\bibfield  {journal} {\bibinfo  {journal} {J Low Temp. Phys.}\ }\textbf {\bibinfo {volume} {110}},\ \bibinfo {pages} {677} (\bibinfo {year} {1998})}\BibitemShut {NoStop}%
\bibitem [{\citenamefont {Kolevski}\ and\ \citenamefont {Boninsegni}(2025)}]{Kolevski2025}%
  \BibitemOpen
  \bibfield  {author} {\bibinfo {author} {\bibfnamefont {K.~M.}\ \bibnamefont {Kolevski}}\ and\ \bibinfo {author} {\bibfnamefont {M.}~\bibnamefont {Boninsegni}},\ }\href {\doibase 10.1063/5.0270881} {\bibfield  {journal} {\bibinfo  {journal} {J. Chem. Phys.}\ }\textbf {\bibinfo {volume} {162}},\ \bibinfo {pages} {174307} (\bibinfo {year} {2025})}\BibitemShut {NoStop}%
\bibitem [{\citenamefont {Nho}\ and\ \citenamefont {Manousakis}(2002)}]{Nho2002}%
  \BibitemOpen
  \bibfield  {author} {\bibinfo {author} {\bibfnamefont {K.}~\bibnamefont {Nho}}\ and\ \bibinfo {author} {\bibfnamefont {E.}~\bibnamefont {Manousakis}},\ }\href@noop {} {\bibfield  {journal} {\bibinfo  {journal} {Phys. Rev. B}\ }\textbf {\bibinfo {volume} {67}},\ \bibinfo {pages} {195411} (\bibinfo {year} {2002})}\BibitemShut {NoStop}%
\bibitem [{\citenamefont {Hu}\ and\ \citenamefont {Boninsegni}(2024)}]{Hu2024}%
  \BibitemOpen
  \bibfield  {author} {\bibinfo {author} {\bibfnamefont {J.-R.}\ \bibnamefont {Hu}}\ and\ \bibinfo {author} {\bibfnamefont {M.}~\bibnamefont {Boninsegni}},\ }\href {\doibase https://doi.org/10.1016/j.rinp.2024.107737} {\bibfield  {journal} {\bibinfo  {journal} {Results Phys.}\ }\textbf {\bibinfo {volume} {61}},\ \bibinfo {pages} {107737} (\bibinfo {year} {2024})}\BibitemShut {NoStop}%
\bibitem [{\citenamefont {Turnbull}\ and\ \citenamefont {Boninsegni}(2005)}]{Turnbull2005}%
  \BibitemOpen
  \bibfield  {author} {\bibinfo {author} {\bibfnamefont {J.~D.}\ \bibnamefont {Turnbull}}\ and\ \bibinfo {author} {\bibfnamefont {M.}~\bibnamefont {Boninsegni}},\ }\href {\doibase 10.1103/PhysRevB.71.205421} {\bibfield  {journal} {\bibinfo  {journal} {Phys. Rev. B}\ }\textbf {\bibinfo {volume} {71}},\ \bibinfo {pages} {205421} (\bibinfo {year} {2005})}\BibitemShut {NoStop}%
\bibitem [{\citenamefont {Hernandez}, \citenamefont {Cole},\ and\ \citenamefont {Boninsegni}(2003)}]{Hernandez2003}%
  \BibitemOpen
  \bibfield  {author} {\bibinfo {author} {\bibfnamefont {E.~S.}\ \bibnamefont {Hernandez}}, \bibinfo {author} {\bibfnamefont {M.~W.}\ \bibnamefont {Cole}}, \ and\ \bibinfo {author} {\bibfnamefont {M.}~\bibnamefont {Boninsegni}},\ }\href {\doibase 10.1103/PhysRevB.68.125418} {\bibfield  {journal} {\bibinfo  {journal} {Phys. Rev. B}\ }\textbf {\bibinfo {volume} {68}},\ \bibinfo {pages} {125418} (\bibinfo {year} {2003})}\BibitemShut {NoStop}%
\bibitem [{\citenamefont {Silvera}\ and\ \citenamefont {Goldman}(1978)}]{Silvera1978}%
  \BibitemOpen
  \bibfield  {author} {\bibinfo {author} {\bibfnamefont {I.~F.}\ \bibnamefont {Silvera}}\ and\ \bibinfo {author} {\bibfnamefont {V.~V.}\ \bibnamefont {Goldman}},\ }\href@noop {} {\bibfield  {journal} {\bibinfo  {journal} {J. Chem. Phys.}\ }\textbf {\bibinfo {volume} {69}},\ \bibinfo {pages} {4209} (\bibinfo {year} {1978})}\BibitemShut {NoStop}%
\bibitem [{\citenamefont {Carlos}\ and\ \citenamefont {Cole}(1980)}]{Carlos1980}%
  \BibitemOpen
  \bibfield  {author} {\bibinfo {author} {\bibfnamefont {W.~E.}\ \bibnamefont {Carlos}}\ and\ \bibinfo {author} {\bibfnamefont {M.~W.}\ \bibnamefont {Cole}},\ }\href@noop {} {\bibfield  {journal} {\bibinfo  {journal} {Surf. Sci.}\ }\textbf {\bibinfo {volume} {91}},\ \bibinfo {pages} {339} (\bibinfo {year} {1980})}\BibitemShut {NoStop}%
\bibitem [{\citenamefont {Boninsegni}, \citenamefont {Prokof'ev},\ and\ \citenamefont {Svistunov}(2006{\natexlab{a}})}]{Boninsegni2006}%
  \BibitemOpen
  \bibfield  {author} {\bibinfo {author} {\bibfnamefont {M.}~\bibnamefont {Boninsegni}}, \bibinfo {author} {\bibfnamefont {N.}~\bibnamefont {Prokof'ev}}, \ and\ \bibinfo {author} {\bibfnamefont {B.}~\bibnamefont {Svistunov}},\ }\href {\doibase 10.1103/PhysRevLett.96.070601} {\bibfield  {journal} {\bibinfo  {journal} {Phys. Rev. Lett.}\ }\textbf {\bibinfo {volume} {96}},\ \bibinfo {pages} {070601} (\bibinfo {year} {2006}{\natexlab{a}})}\BibitemShut {NoStop}%
\bibitem [{\citenamefont {Boninsegni}, \citenamefont {Prokof'ev},\ and\ \citenamefont {Svistunov}(2006{\natexlab{b}})}]{Boninsegni2006b}%
  \BibitemOpen
  \bibfield  {author} {\bibinfo {author} {\bibfnamefont {M.}~\bibnamefont {Boninsegni}}, \bibinfo {author} {\bibfnamefont {N.~V.}\ \bibnamefont {Prokof'ev}}, \ and\ \bibinfo {author} {\bibfnamefont {B.~V.}\ \bibnamefont {Svistunov}},\ }\href {\doibase 10.1103/PhysRevE.74.036701} {\bibfield  {journal} {\bibinfo  {journal} {Phys. Rev. E}\ }\textbf {\bibinfo {volume} {74}},\ \bibinfo {pages} {036701} (\bibinfo {year} {2006}{\natexlab{b}})}\BibitemShut {NoStop}%
\bibitem [{\citenamefont {Feynman}\ and\ \citenamefont {Hibbs}(1965)}]{Feynman1965}%
  \BibitemOpen
  \bibfield  {author} {\bibinfo {author} {\bibfnamefont {R.~P.}\ \bibnamefont {Feynman}}\ and\ \bibinfo {author} {\bibfnamefont {A.~R.}\ \bibnamefont {Hibbs}},\ }\enquote {\bibinfo {title} {{Quantum Mechanics} {and Path Integrals}},}\ \ (\bibinfo  {publisher} {McGraw-Hill, New York},\ \bibinfo {year} {1965})\ Chap.~\bibinfo {chapter} {10}\BibitemShut {NoStop}%
\bibitem [{\citenamefont {Pollock}\ and\ \citenamefont {Ceperley}(1989)}]{Pollock1987}%
  \BibitemOpen
  \bibfield  {author} {\bibinfo {author} {\bibfnamefont {E.~L.}\ \bibnamefont {Pollock}}\ and\ \bibinfo {author} {\bibfnamefont {D.~M.}\ \bibnamefont {Ceperley}},\ }\href {\doibase 10.1103/PhysRevB.36.8343} {\bibfield  {journal} {\bibinfo  {journal} {Phys. Rev. B}\ }\textbf {\bibinfo {volume} {36}},\ \bibinfo {pages} {8343} (\bibinfo {year} {1989})}\BibitemShut {NoStop}%
\bibitem [{\citenamefont {Sindzingre}, \citenamefont {Klein},\ and\ \citenamefont {Ceperley}(1989)}]{Sindzingre1989}%
  \BibitemOpen
  \bibfield  {author} {\bibinfo {author} {\bibfnamefont {P.}~\bibnamefont {Sindzingre}}, \bibinfo {author} {\bibfnamefont {M.~L.}\ \bibnamefont {Klein}}, \ and\ \bibinfo {author} {\bibfnamefont {D.~M.}\ \bibnamefont {Ceperley}},\ }\href {\doibase 10.1103/physrevlett.63.1601} {\bibfield  {journal} {\bibinfo  {journal} {Phys. Rev. Lett.}\ }\textbf {\bibinfo {volume} {63}},\ \bibinfo {pages} {1601} (\bibinfo {year} {1989})}\BibitemShut {NoStop}%
\bibitem [{\citenamefont {Boninsegni}\ and\ \citenamefont {Ceperley}(1995)}]{Boninsegni1995}%
  \BibitemOpen
  \bibfield  {author} {\bibinfo {author} {\bibfnamefont {M.}~\bibnamefont {Boninsegni}}\ and\ \bibinfo {author} {\bibfnamefont {D.~M.}\ \bibnamefont {Ceperley}},\ }\href {\doibase 10.1103/PhysRevLett.74.2288} {\bibfield  {journal} {\bibinfo  {journal} {Phys. Rev. Lett.}\ }\textbf {\bibinfo {volume} {74}},\ \bibinfo {pages} {2288} (\bibinfo {year} {1995})}\BibitemShut {NoStop}%
\bibitem [{\citenamefont {Boninsegni}\ \emph {et~al.}(2012)\citenamefont {Boninsegni}, \citenamefont {Pollet}, \citenamefont {Prokof'ev},\ and\ \citenamefont {Svistunov}}]{Boninsegni2012}%
  \BibitemOpen
  \bibfield  {author} {\bibinfo {author} {\bibfnamefont {M.}~\bibnamefont {Boninsegni}}, \bibinfo {author} {\bibfnamefont {L.}~\bibnamefont {Pollet}}, \bibinfo {author} {\bibfnamefont {N.}~\bibnamefont {Prokof'ev}}, \ and\ \bibinfo {author} {\bibfnamefont {B.}~\bibnamefont {Svistunov}},\ }\href {\doibase 10.1103/PhysRevLett.109.025302} {\bibfield  {journal} {\bibinfo  {journal} {Phys. Rev. Lett.}\ }\textbf {\bibinfo {volume} {109}},\ \bibinfo {pages} {025302} (\bibinfo {year} {2012})}\BibitemShut {NoStop}%
\bibitem [{Note1()}]{Note1}%
  \BibitemOpen
  \bibinfo {note} {An illustrative animation of representative imaginary-time many-particle paths for a ($p$-H$_2$)$_8$ cluster at $T=0.125$ K can be viewed at \protect \url {https://doi.org/10.5281/zenodo.16808901}.}\BibitemShut {Stop}%
\bibitem [{\citenamefont {Bretz}\ \emph {et~al.}(1973)\citenamefont {Bretz}, \citenamefont {Dash}, \citenamefont {Hickernell}, \citenamefont {McLean},\ and\ \citenamefont {Vilches}}]{Bretz1973}%
  \BibitemOpen
  \bibfield  {author} {\bibinfo {author} {\bibfnamefont {M.}~\bibnamefont {Bretz}}, \bibinfo {author} {\bibfnamefont {J.~G.}\ \bibnamefont {Dash}}, \bibinfo {author} {\bibfnamefont {D.~C.}\ \bibnamefont {Hickernell}}, \bibinfo {author} {\bibfnamefont {E.~O.}\ \bibnamefont {McLean}}, \ and\ \bibinfo {author} {\bibfnamefont {O.~E.}\ \bibnamefont {Vilches}},\ }\href {\doibase 10.1103/PhysRevA.8.1589} {\bibfield  {journal} {\bibinfo  {journal} {Phys. Rev. A}\ }\textbf {\bibinfo {volume} {8}},\ \bibinfo {pages} {1589} (\bibinfo {year} {1973})}\BibitemShut {NoStop}%
\bibitem [{\citenamefont {Nielsen}, \citenamefont {McTague},\ and\ \citenamefont {Passell}(1980)}]{Nielsen1980}%
  \BibitemOpen
  \bibfield  {author} {\bibinfo {author} {\bibfnamefont {M.}~\bibnamefont {Nielsen}}, \bibinfo {author} {\bibfnamefont {J.~P.}\ \bibnamefont {McTague}}, \ and\ \bibinfo {author} {\bibfnamefont {L.}~\bibnamefont {Passell}},\ }in\ \href {\doibase 10.1007/978-1-4613-3057-8_5} {\emph {\bibinfo {booktitle} {{Phase Transitions} {in Surface Films}}}},\ \bibinfo {editor} {edited by\ \bibinfo {editor} {\bibfnamefont {J.~G.}\ \bibnamefont {Dash}}\ and\ \bibinfo {editor} {\bibfnamefont {J.}~\bibnamefont {Ruvalds}}}\ (\bibinfo  {publisher} {Springer US},\ \bibinfo {address} {Boston, MA},\ \bibinfo {year} {1980})\ pp.\ \bibinfo {pages} {127--163}\BibitemShut {NoStop}%
\bibitem [{\citenamefont {Greywall}(1993)}]{Greywall1993}%
  \BibitemOpen
  \bibfield  {author} {\bibinfo {author} {\bibfnamefont {D.~S.}\ \bibnamefont {Greywall}},\ }\href {\doibase 10.1103/PhysRevB.47.309} {\bibfield  {journal} {\bibinfo  {journal} {Phys. Rev. B}\ }\textbf {\bibinfo {volume} {47}},\ \bibinfo {pages} {309} (\bibinfo {year} {1993})}\BibitemShut {NoStop}%
\bibitem [{\citenamefont {Boninsegni}(2012)}]{Boninsegni2012b}%
  \BibitemOpen
  \bibfield  {author} {\bibinfo {author} {\bibfnamefont {M.}~\bibnamefont {Boninsegni}},\ }\href@noop {} {\bibfield  {journal} {\bibinfo  {journal} {J. Low. Temp. Phys.}\ }\textbf {\bibinfo {volume} {168}},\ \bibinfo {pages} {137} (\bibinfo {year} {2012})}\BibitemShut {NoStop}%
\bibitem [{\citenamefont {Kora}\ and\ \citenamefont {Boninsegni}(2019)}]{Kora2019}%
  \BibitemOpen
  \bibfield  {author} {\bibinfo {author} {\bibfnamefont {Y.}~\bibnamefont {Kora}}\ and\ \bibinfo {author} {\bibfnamefont {M.}~\bibnamefont {Boninsegni}},\ }\href {\doibase 10.1007/s10909-019-02229-z} {\bibfield  {journal} {\bibinfo  {journal} {J. Low. Temp. Phys.}\ }\textbf {\bibinfo {volume} {197}},\ \bibinfo {pages} {337} (\bibinfo {year} {2019})}\BibitemShut {NoStop}%
\bibitem [{\citenamefont {Deltour}, \citenamefont {Barrat},\ and\ \citenamefont {Jensen}(1997)}]{Deltour1997}%
  \BibitemOpen
  \bibfield  {author} {\bibinfo {author} {\bibfnamefont {P.}~\bibnamefont {Deltour}}, \bibinfo {author} {\bibfnamefont {J.-L.}\ \bibnamefont {Barrat}}, \ and\ \bibinfo {author} {\bibfnamefont {P.}~\bibnamefont {Jensen}},\ }\href {\doibase 10.1103/PhysRevLett.78.4597} {\bibfield  {journal} {\bibinfo  {journal} {Phys. Rev. Lett.}\ }\textbf {\bibinfo {volume} {78}},\ \bibinfo {pages} {4597} (\bibinfo {year} {1997})}\BibitemShut {NoStop}%
\bibitem [{\citenamefont {Ala-Nissila}, \citenamefont {Ferrando},\ and\ \citenamefont {Ying}(2002)}]{Ala-Nissila2002}%
  \BibitemOpen
  \bibfield  {author} {\bibinfo {author} {\bibfnamefont {T.}~\bibnamefont {Ala-Nissila}}, \bibinfo {author} {\bibfnamefont {R.}~\bibnamefont {Ferrando}}, \ and\ \bibinfo {author} {\bibfnamefont {S.~C.}\ \bibnamefont {Ying}},\ }\href {\doibase 10.1080/00018730110107902} {\bibfield  {journal} {\bibinfo  {journal} {Adv. Phys.}\ }\textbf {\bibinfo {volume} {51}},\ \bibinfo {pages} {949} (\bibinfo {year} {2002})}\BibitemShut {NoStop}%
\bibitem [{\citenamefont {Hornek\ae{}r}\ \emph {et~al.}(2006)\citenamefont {Hornek\ae{}r}, \citenamefont {Rauls}, \citenamefont {Xu}, \citenamefont {{\u{S}}ljivan{\u{c}}anin}, \citenamefont {Otero}, \citenamefont {Stensgaard}, \citenamefont {L\ae{}gsgaard}, \citenamefont {Hammer},\ and\ \citenamefont {Besenbacher}}]{Hornekær2006}%
  \BibitemOpen
  \bibfield  {author} {\bibinfo {author} {\bibfnamefont {L.}~\bibnamefont {Hornek\ae{}r}}, \bibinfo {author} {\bibfnamefont {E.}~\bibnamefont {Rauls}}, \bibinfo {author} {\bibfnamefont {W.}~\bibnamefont {Xu}}, \bibinfo {author} {\bibfnamefont {{\u{Z}}.}~\bibnamefont {{\u{S}}ljivan{\u{c}}anin}}, \bibinfo {author} {\bibfnamefont {R.}~\bibnamefont {Otero}}, \bibinfo {author} {\bibfnamefont {I.}~\bibnamefont {Stensgaard}}, \bibinfo {author} {\bibfnamefont {E.}~\bibnamefont {L\ae{}gsgaard}}, \bibinfo {author} {\bibfnamefont {B.}~\bibnamefont {Hammer}}, \ and\ \bibinfo {author} {\bibfnamefont {F.}~\bibnamefont {Besenbacher}},\ }\href {\doibase 10.1103/PhysRevLett.97.186102} {\bibfield  {journal} {\bibinfo  {journal} {Phys. Rev. Lett.}\ }\textbf {\bibinfo {volume} {97}},\ \bibinfo {pages} {186102} (\bibinfo {year} {2006})}\BibitemShut {NoStop}%
\end{thebibliography}%
\end{document}